\documentclass[twocolumn]{aastex631}
\usepackage{enumitem}
\usepackage{bm}
\usepackage{comment}
\usepackage{graphicx}
\graphicspath{{./figures/}{./}}
\usepackage{hyperref}
\usepackage{soul} 
\usepackage{amsmath}
\usepackage{amssymb}
\usepackage{xspace}
\usepackage{xifthen}

\newcommand{\ie}{i.e.\xspace}
\newcommand{\eg}{e.g.\xspace}

\newcommand{\vs}{vs.\xspace}

\mathchardef\mhyphen="2D

\newlength{\dhatheight}

\newcommand{\unit}[1]{\ensuremath{\mathrm{\,#1}}\xspace}
\newcommand{\Myr}{\unit{Myr}}
\newcommand{\Gyr}{\unit{Gyr}}

\newcommand{\km}{\unit{km}}
\newcommand{\kms}{\km \second^{-1}}

\newcommand{\kpc}{\unit{kpc}}

\newcommand{\second}{\unit{s}}

\newcommand{\Msun}{\unit{M_\odot}}

\newcommand{\e}{\unit{e^{-}}}

\newcommand{\kpckms}{\kpc \km \second^{-1}}

\newcommand{\secref}[1]{Section~\ref{sec:#1}}

\newcommand{\figref}[1]{Fig.~\ref{fig:#1}}

\newcommand{\bandvar}[2][]{%
  \ifthenelse{\isempty{#1}}{\var{#2}}{\var{#2\_#1}}%
}

\newcommand{\Gaia}{\textit{Gaia}\xspace}

\newcommand{\var}[1]{\ensuremath{\texttt{\MakeUppercase{#1}}}\xspace}

\providecommand\physrep{\ref@jnl{Phys.~Rep.}}%
\providecommand\apjs{\ref@jnl{ApJS}}%
\providecommand{\jcap}{\ref@jnl{JCAP}}%

\usepackage[absolute,overlay]{textpos}

\begin{document}

\shortauthors{Tavangar et al.}

\title{Phase spirals across galactic disks I: Exploring dynamical influences on winding}

\author[0000-0001-6584-6144]{Kiyan Tavangar}
\affiliation{Department of Astronomy, Columbia University, New York, NY 10027, USA}

\author[0000-0001-6244-6727]{Kathryn V. Johnston}
\affiliation{Department of Astronomy, Columbia University, New York, NY 10027, USA}

\author[0000-0001-8917-1532]{Jason A. S. Hunt}
\affiliation{School of Mathematics \& Physics, University of Surrey, \\Stag Hill, Guildford, GU2 7XH, UK}

\author[0000-0001-5686-3743]{Axel Widmark}
\affiliation{Department of Astronomy, Columbia University, New York, NY 10027, USA}
\affiliation{Stockholm University and The Oskar Klein Centre for Cosmoparticle Physics, \\
Alba Nova, 10691 Stockholm, Sweden}

\author[0000-0002-5861-5687]{Chris Hamilton}
\affiliation{School of Natural Sciences, Institute for Advanced Study, Princeton, NJ 08540, USA}
\affiliation{Department of Astrophysical Sciences, 4 Ivy Lane, Princeton University, Princeton, NJ 08544, USA}

\author[0000-0003-1517-3935]{Michael S. Petersen}
\affiliation{Institute for Astronomy, University of Edinburgh, Royal Observatory, Blackford Hill, Edinburgh EH9 3HJ, UK}

\author[0000-0003-2660-2889]{Martin D. Weinberg}
\affiliation{Department of Astronomy, University of Massachusetts at Amherst, 710 N. Pleasant St., Amherst, MA 01003}


\correspondingauthor{Kiyan Tavangar}
\email{k.tavangar@columbia.edu}

\begin{abstract}
The vertical phase-space spirals in the Milky Way are clear evidence of disequilibrium.
However, they are challenging to study because phase mixing signals evolve under the influence of many different dynamical processes and can be driven by many sources of disequilibrium.
We characterize phase spirals in two simulations -- one test particle and one N-body -- with basis function expansions, using these to derive winding times ($T_{\rm fit}$).
We find that phase spirals in the test particle simulation wind up 
as expected from pure phase mixing theory while those in the self-consistent simulation do not. 
Specifically, in the N-body simulation we find that (i) the onset of winding is delayed, (ii) the winding rate is slowed, and (iii) the rate of winding oscillates with time.
The extent of these effects depends on the azimuthal action $J_\phi$ of the phase spiral region.
We build some physical intuition for these effects through 1-D toy models which follow a group of co-moving stars traveling through several different evolving potentials.
We find that phase spiral winding can be delayed until the group no longer moves coherently with the midplane of the (perturbed) potential 
and oscillates with time as the group experiences (e.g.) a breathing mode traveling through the disk.
The modifications to winding are strongest in the inner galaxy where the disk potential dominates.
We conclude that in the Milky Way, all calculations of the winding time should be interpreted as lower limits and that the most trustworthy winding times are likely in the outer disk.
\end{abstract}


\section{Introduction} \label{sec:intro}

Historically, the Milky Way (MW) disk has been assumed to be in approximate equilibrium.
This allowed researchers to simplify the exploration of the Galaxy's different components and generate broad conclusions about its formation and evolution \citep{bland2016galaxy}.
However, in the past decade the \Gaia space telescope \citep{Gaia:2016} has provided an unprecedentedly detailed view of the Galactic disk \citep[see \eg][and references therein]{Hunt&Vasiliev:2025}.
The resulting dataset revealed new signatures of disk disequilibrium and allowed detailed exploration of these structures, bringing us closer to a complete picture of the Galaxy.

One prominent example of this is that, using \Gaia DR2 data, \citet{Antoja:2018} found a spiral structure in the ($z,v_z$) density of Solar Neighborhood stars (within $0.1$ \kpc of the Sun).
This spiral feature has since become known as the vertical phase spiral (the terminology we will use throughout this work) or the \Gaia ``snail'', and is a clear contradiction of our assumption of equilibrium in our Galaxy.
This expectation for an equilibrium disk can be understood by considering how vertical oscillations of stars around the Galactic midplane project into the ($z,v_z$) plane.
In this space, the oscillations correspond to rotations around $(z, v_z) = (0,0)$, following (nearly) elliptical paths.
The angles between star locations and the positive $z$-axis characterize the phase of the oscillations (proxies for vertical angles $\theta_z$), while the area within these ellipses characterize the vertical energy \citep[proxies for  vertical actions, $J_z$,  see, \eg][]{Price-Whelan:2021, Horta:2024}.
When the disk is in equilibrium, stars in any given disk region should be evenly distributed in phase and follow some smooth density profile in vertical energy.
The distribution would then resemble a featureless monopole, axisymmetric about $(z, v_z) = (0,0)$.

The Solar Neighborhood phase spiral is present not only in density but also when weighting by Galactocentric radial velocity, azimuthal velocity, age, and even chemistry \citep[\eg][]{Antoja:2018, Bland-Hawthorn:2019, Frankel:2025}.
Furthermore, various studies have now shown that other parts of the Galactic disk also have phase spirals \citep{Bland-Hawthorn:2019, Li:2021, Hunt:2022}.
The phase spirals in different regions of the disk vary in morphology, including the amplitude of the spiral feature relative to the equilibrium background, how wound up the spiral is, and the number of spiral arms.

There are three basic ingredients required to create a phase spiral: (i) a perturbation causing an asymmetry in the vertical oscillation phase, (ii) a gradient with orbital properties in the quantity being visualized in the $(z,v_z)$ plane (e.g. decreasing density with scale height of populations), and (iii) an anharmonic vertical potential, in which oscillation frequencies vary as a function of energy \citep[\eg][]{Banik:2022, Banik:2023}. 
Consider as an example an impulsive perturbation which imparts a small positive velocity kick to all stars in a small patch of the disk.
All stars in this region, originally in an equilibrium monopole distribution, will simultaneously move along the $v_z$ axis in $(z,v_z)$ space.
As a result, there will be an asymmetry in vertical oscillation phase, specifically an overdensity of stars moving with positive $v_z$ and an underdensity moving with negative $v_z$.
This newly perturbed distribution is a dipole, which can be thought of as a completely unwound spiral.
To create a recognizable spiral feature like that seen in the data, the dipole needs to wind up.
This will occur in any anharmonic vertical potential in which stars with lower vertical energies (\ie those with low maximum $z$ on their orbits) oscillate with higher vertical frequencies.
In $(z,v_z)$ space, this translates as stars with lower vertical energies completing an elliptical rotation faster.
This means that a set of stars at equivalent vertical orbital phase (\ie a line extending from the origin in $(z,v_z)$ space) will spread out into a spiral in $(z,v_z)$ over time.
In our example of the impulsive perturbation, the dipole will therefore wind up into a one-armed spiral.
Two-armed spirals can be formed similarly if the initial perturbation is quadrupolar rather than dipolar.
In this simple model of an impulsive perturbation in a known potential, one can calculate the frequencies of stars, one can calculate the winding rate of the phase spiral and predict its evolution over time.
Conversely, a fully evolved spiral can be `rewound' back to its initial dipole, and hence the time of the perturbation can be derived.
In the real MW, however, there are various reasons why this simple picture does not work.

First, the perturbation which caused the phase spiral in the MW is not known.
Some possibilities include satellites such as the Sagittarius dwarf galaxy \citep[\eg][]{Antoja:2018, Binney&Schonrich:2018, Darling:19a, Laporte:2019, Bland-Hawthorn:2019, Hunt:2021, Gandhi:2022, Darragh-Ford:2023}, the dark matter wakes they induce \citep{Grand:2023}, the buckling of the Galactic bar \citep{Khoperskov:2019}, the spiral arms \citep{Faure:2014, Hunt:2022, Li:2023}, a combination of large and small-scale kicks from substructure in the global potential \citep{Tremaine:2023, Gilman:2025}, and ``galactic echoes'' from the nonlinear coupling of two successive short-lived perturbations \citep{Chiba:2025}.
However, none of these can individually explain the observed phase spirals \citep[\eg][]{Laporte:2019, Bennett&Bovy:2021, Quillen:2018}.
In reality, all of these processes are active in the Galaxy, each forming phase-spiral-like features at different rates, amplitudes, and locations across the disk.

Second, recent analyses have shown that the MW phase spirals are unlikely to evolve precisely as predicted from pure phase mixing theory.
This is because in self-consistent simulations, following a perturbation the phase spiral winding is often found to be initially delayed and then to proceed more slowly than naively expected \citep{Darling:19a, Darling:19b, Darling:21, Bennett&Bovy:2021, Widrow:2023, Bland-Hawthorn_Tepper-Garcia:2021, Darling:24, Asano&Antoja:2025}.
Using a high resolution N-body simulation modeled after the MW, \citet{Asano&Antoja:2025} suggest that the winding delay in the MW is $\simeq 300$ \Myr, meaning previous estimates of the winding time based purely on phase mixing theory \citep[$0.2-1$ \Gyr;][]{Antoja:2018, Li&Shen:2020, Li:2021, Widmark:2022, Frankel:2023, Darragh-Ford:2023, Antoja:2023, Frankel:2025, Widmark:2025, Hunt&Vasiliev:2025} are likely significant underestimates.

A few recent theoretical studies have aimed at understanding the physics of self-gravitating vertical disk oscillations.
For instance, \citet{Widrow:2023} studies the phase spirals generated when one perturbs a self-gravitating 3D shearing box, which is supposed to mimic a patch of our Galactic disk.
He found that swing-amplified perturbations close to the disk plane acted to drive stronger and more slowly-winding vertical phase spirals compared to the test particle case. 
However, his analysis was limited by the fact that the center of the 3D box had to be pinned to the midplane of the galaxy, and that the analysis was necessarily local rather than global.
By contrast, \citet{Binney:2024} examined global self-gravitating disk distortions driven by a passing satellite, allowing the entire disk plane to be `plucked' self-consistently without pinning the coordinate system. 
He concluded that one must analyze the entire system together (\ie disk and satellite) in this way to uncover the underlying dynamics. However, to make his calculation tractable \citet{Binney:2024} had to treat his disk as a pressureless fluid, ignoring all velocity dispersion, and so he was not able to study phase spirals. 

Our approach combines some of the merits of these two papers: like \citet{Widrow:2023} we compare test particle and self-consistent phase-spiral dynamics, but like \citet{Binney:2024} we work in a global setting in which the entire disk responds to a satellite. We focus in particular on how the winding behavior of phase spirals is modified in the self-consistent case versus the test particle case, and especially how this modification depends on orbital angular momentum $J_\phi$.

This paper is organized as follows.
In \secref{sims} we present two simulations analyzed and compared in this work: a test particle simulation and an N-body one.
In \secref{methods}, we explain the different analysis techniques we use to explore phase spiral evolution, including our choice of coordinates and how we use basis function expansions to characterize the phase spirals.
In \secref{results}, we present the derived winding times for phase spirals across the disk in both simulations, interpret and explain those results using physical intuition and a toy model, and connect our results to observations.
In \secref{discussion}, we compare our results to previous studies. We conclude in \secref{conclusion}.
In a companion paper, we will build on the work here by using the basis function expansion technique to analyze correlations between phase spiral properties across the face of our simulated galactic disk.

\section{Simulations} \label{sec:sims}

\begin{figure*}[th!]
    \centering
    \includegraphics[width=\linewidth]{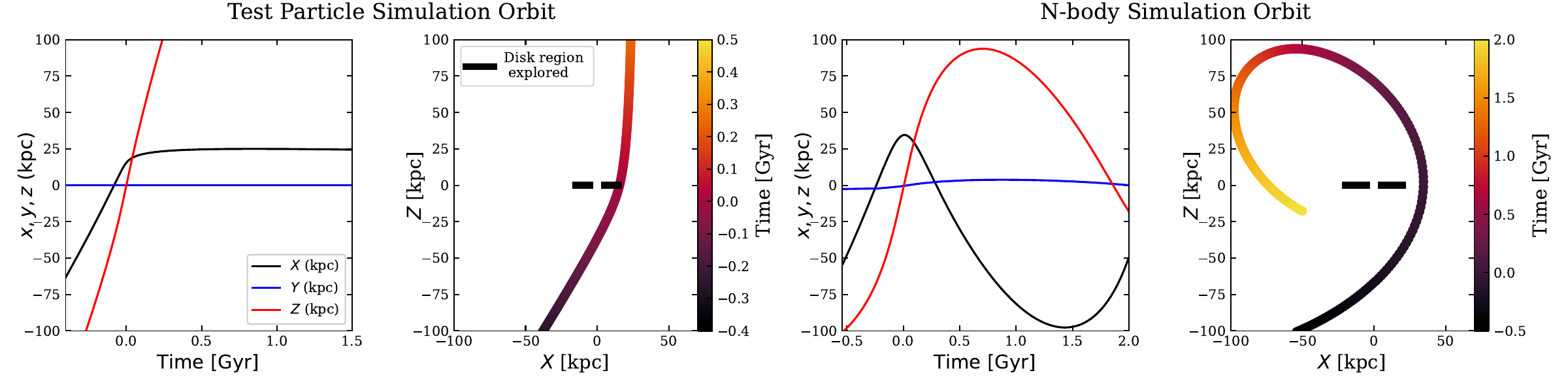}
    \caption{The orbit of the perturber in both the test particle (left panels) and N-body (right panels) simulations. $T=0$ in these panels corresponds to the time of the disk crossing.}
    \label{fig:sim_orbits}
\end{figure*}

We use both a test particle and a fully self-consistent N-body simulation to examine phase spiral winding following a perturbation by a satellite.
Besides the introduction of N-body interactions in the latter case, these two simulations are qualitatively similar.
The two disks contain $2\times10^8$ and $2.2\times10^8$ particles, respectively.
These therefore pass the lower threshold of $10^8$ particles suggested by \citet{Binney&Schonrich:2018} to ensure sufficient phase space resolution. 
They both include a satellite whose internal properties are designed to approximate that of the Sagittarius dwarf galaxy (Sgr), one of the most likely culprits for the MW phase spirals.
However, our purpose is not to rule out or confirm Sgr as the origin of the phase spirals in the MW.
Rather, we will use a comparison of the phase spiral evolution between the two simulations to isolate how phase spirals wind differently in self-consistent simulations versus test-particle ones. 

\subsection{Test Particle Simulation} \label{sec:test_sim}

Our test particle simulation contains a MW-like host and a dwarf galaxy satellite.
The structures of both galaxies are represented by static, analytic functions.
For the host galaxy, we use the \texttt{MWPotential2014} from the Galactic dynamics package \texttt{galpy} \citep{Bovy:2015}.
This model has three components for the disk, bulge, and halo, and we refer the interested reader to Table 1 of \citet{Bovy:2015} for more details about each.

For the satellite, we use a Plummer potential \citep{Plummer:11}, with mass $M=3\times10^9 \Msun$ and scale radius $b=1 \kpc$. 
We initialize this satellite at position $(x,y,z) = (-63, 0,-140) \kpc$ with velocity $(v_x,v_y,v_z) = (180, 0,282) \kms$ and show its resulting orbit in the two left-hand panels of \figref{sim_orbits}.
It is on an unbound orbit with a pericenter of $\simeq 15 \kpc$.
It completes a single fly-by, crossing the disk almost perpendicularly with velocity $\simeq 500 \kms$.
This disk crossing occurs 0.4 \Gyr after the start of the simulation when the satellite is at pericenter.
To simplify our analysis, we redefine $T=0$ to be the disk crossing time.
In the actual MW or a self-consistent simulation, the satellite's passage would lead to a reflex motion from the MW disk away from its initial flat distribution centered on $z=0$. 
In this test particle simulation though, we fix the MW potential to (0,0) for simplicity which means we do not account for any non-inertial effects from the mass movement of the disk.
However, given the relatively impulsive nature of this simulation and its perpendicular trajectory through the disk, we would not expect significant movement from the disk as a whole.

Finally, we note, this orbit is quite different from the true Sgr orbit, which is bound to the MW and has completed a few disk crossings. 
It was instead chosen so we could study the response of the disk to a single encounter.

We then run the simulation for $3 \Gyr$.


\subsection{Self-consistent N-body simulation} \label{sec:nbody_sim}
To model a self-gravitating MW--Sgr analog, we use the M1 model\footnote{https://www.sciserver.org/datasets/cosmology/smudge/} from \citet{Hunt:2021}.
The M1 model is a high resolution ($> 10^9$ particle) simulation evolved with a GPU accelerated N-body code called \textsc{Bonsai} \citep{Bedorf:2012}.

The host galaxy in the M1 simulation is a disk-bulge-halo model (MWb in \citet{Widrow&Dubinski:2005}).
The disk is assumed to be axisymmetric with a quasi--Maxwellian distribution function taken from \citet{Kuijken&Dubinski:1995}. 
Its radial surface density profile is exponential with a scale length of $2.817$ \kpc and the vertical structure is given by an anharmonic sech$^2$ density profile (\ie an $\ln \cosh$ potential profile) with scale height $0.439$ \kpc.
The bulge is a Hernquist model \citep{Hernquist:1990} with scale length 0.884 \kpc.
The dark halo takes a Navarro-Frank-White profile \citep{NFW:1997} with mass $\simeq 6\times10^{11} \Msun$ and scale radius 8.818 \kpc.
More details about the initial parameters are given in Table 2 of \citet{Widrow&Dubinski:2005}.

\citet{Hunt:2021} initialized the disk, bulge, and halo particles using the parallelized version of the \textsc{galactics}\footnote{https://github.com/treecode/galactics.parallel} \citep{Kuijken&Dubinski:1995} initial condition generator.
The dwarf galaxy initial conditions are taken from model L2 of \citet{Laporte:2018}, which is composed of two Hernquist spheres \citep{Hernquist:1990}.
The first represents the dark matter, with virial mass $M_{200}=6 \times 10^{10}  M_{\odot}$, concentration parameter $c_{200}=28$, halo mass $M_h = 8 \times 10^{10} M_{\odot}$, and scale radius $a_h=8$ \kpc. 
The second represents the stellar component within the dark matter halo with $M_*=6.4 \times 10^8 M_{\odot}$ and $a_h=0.85$ \kpc.
For a more detailed description of the satellite, see \citet{Laporte:2018}.

The simulation is run for $\simeq 8.3 \Gyr$, with force calculations every 9.778 thousand years and snapshot outputs every 9.778 \Myr.
In this simulation, the satellite is bound to the host, meaning it completes multiple orbits and disk crossings.
It is initialized approximately at apocenter at position $(x,y,z) = (-244, 0,-90) \kpc$ with velocity $(v_x,v_y,v_z) = (43, -2.0,-37) \kms$.
The first two disk crossings within $50 \kpc$ occur $\simeq 2.5$ and $4.8$ \Gyr into the simulation at $\simeq 350 \kms$ and $\simeq 275 \kms$.
In between these two disk passages phase spirals evolve from a single perturbation, making this period qualitatively similar to the test particle case.
As a result, in an effort to ease comparisons with the test particle simulation, we focus our analysis of the disk on the 2 \Gyr after the first disk passage.
We choose to define $T=0$ as the time of the first disk crossing, which occurs at pericenter at $\simeq 35$ \kpc from the host galaxy center.
The orbit of the satellite in this time interval is shown in the right two panels of \figref{sim_orbits}.
We leave exploration of how multiple passages can affect phase spiral formation and evolution to future exploration.

Apart from the different satellite orbits, we note two additional differences between the test particle and N-body models.
First and most importantly, the latter is a self-consistent simulation.
Second, while both disk potentials seek to approximate the MW potential, they do have slightly different density profiles.
Specifically, the N-body disk is slightly thicker and less centrally concentrated than the test particle disk.
As a result, it has more stars in the outer regions of the disk, meaning we are able to examine phase spirals out to greater Galactocentric radii.
Despite the difference in potentials, we believe it is still appropriate to compare these two simulations to analyze how incorporating N-body interactions 
affects phase spiral formation and evolution.
The test particle case provides a clear demonstration of the validity of our intuition for the evolution of the phase-spiral due to phase-mixing alone after an isolated satellite interaction.
It is true that the precise shape of the equilibrium potential will have a quantitative influence on the rate of phase spiral winding, and that different interactions (from a single satellite or other perturbations)  will provide different initial conditions that instigate that evolution.
However, our purpose is to study the deviation of the N-body case from our expectations for phase-mixing, which we can do qualitatively, without a direct quantitative comparison of the simulations. 

\section{Analysis Methods} \label{sec:methods}

This section describes our methods for analyzing each simulation snapshot.
First we choose our coordinate systems (\secref{coords}), then bin the data (\secref{bins}), and finally quantify the properties of the phase spirals within each bin (\secref{snails}).

\subsection{Choice of coordinate systems} \label{sec:coords}

Throughout our analysis, we use angle-action rather than physical coordinates.
Actions ($\bm{J}$), which have the same units as angular momentum, are integrals of motion which are constant along an unperturbed orbit, while angles ($\bm{\theta}$) describe the phase along the orbit and are traversed with constant frequency [$\bm{\Omega}(\bm{J})$].
We use \texttt{Galpy} \citep{Bovy:2015} in the test particle case and \texttt{Agama} \citep{Vasiliev:19} in the N-body case to calculate each particle's actions, angles, and frequencies from their 6-D phase space positions and the underlying potential -- in the N-body case derived from the simulation at each timestep.


\subsubsection{Using $(\theta_z, J_z)$ to simplify phase spiral morphology}
\label{sec:actions}

We project from $(z,v_z)$ to $(\theta_z,J_z)$ coordinates. 
This step is motivated by considerations of particle orbits.
In $(z,v_z)$ physical space, stars on near-circular orbits {\it approximately} trace ellipses.
The area enclosed by the ellipse gives a {\it rough estimate} of $J_z$ and the position along the ellipse can be used to indicate $\theta_z$ \citep[e.g.][]{Price-Whelan:2021}. 
While this comparison is conceptually simple, the italicized differences make direct dynamical interpretations of distributions in $(z,v_z)$ space challenging. 
First, the orbits in $(z,v_z)$ are in general not closed, so the path is not strictly repeated.
Second, the progression in time along that path is non-uniform. 
Third, the shape of the path, while simple, deviates significantly from something that can be described with compact formulae.
Finally, orbits in this coordinate system typically overlap with each other, unless restricting to those with zero radial action (or, equivalently, eccentricity). 
Combined, these lead to distributions (and phase spirals) in $(z,v_z)$ that are, at certain radii, non-circular (\ie more ``angular'' or pointy) and non-trivially represented.

Distributions in  $(z,v_z)$ can directly be compared to those in $(\theta_z,J_z)$ by plotting the angle-action variables in polar coordinates $(\sqrt{J_z}\cos{\theta_z},\sqrt{J_z}\sin{\theta_z})$. 
In this space, star orbits follow (exact and closed) circles of radius $\sqrt{J_z}$, with phases increasing steadily from position $\theta_{z,0}$ with time $t$ as $\theta_z = \Omega _z t + \theta_{z,0}$. 
After a perturbation, these distributions spiral outward steadily as the stars within them follow a progression of circular paths. 

For reference, the actions $J_\phi$ and $J_z$ are related to guiding radii $R_G$ and maximum vertical oscillation amplitude $z_{\rm max}$ via

\begin{equation}
R_G \equiv \frac{J_{\phi}} {v_{\rm circ}(J_\phi)} \simeq 8.2 \: {\rm kpc} \frac{J_\phi}{1900 \: {\rm kpc\; km/s}} \frac{230 \:{\rm km/s}}{v_{\rm circ}}
\end{equation}
\begin{equation}
z_{\rm max} \simeq \frac{2J_{z}} {v_{z,{\rm max}}(J_z)} \simeq 300 \: {\rm pc} \frac{2J_z} {9 \:{\rm kpc\; km/s}} \frac{30 \: {\rm km/s}} {v_{z,{\rm max}}}
\end{equation}
In these equations, $v_{\rm circ}(J_\phi)$ is the circular velocity of a star with a given $J_\phi$, $v_{z,{\rm max}}(J_z)$ is the maximum vertical velocity of a star with a given $J_z$, and the numbers on the right are approximations for the Solar Neighborhood.

\subsubsection{Binning particles with a common history}
\label{sec:bins}

Working in angle-action space rather than physical space also ensures we can group stars together that have similar orbital frequencies and, likely, common histories.
Consider two stars that are at some point in close physical proximity (\ie nearby in ($x,y$) coordinates), but with one at its orbital apocenter and one at pericenter.
The star at pericenter (apocenter) will, on average, orbit the galaxy at a smaller (larger) radius, meaning it will have a shorter (longer) orbital period.
Therefore, these two stars will not have been close to one another for the overwhelming majority of their orbital histories. 
As a consequence, they will experience different effects from both the underlying galactic potential and any perturbative forces. If we instead select two disk stars for their proximity in angle-action space, with small $(\delta \theta_\phi,\delta J_\phi)$\footnote{$\Omega_\phi$ depends on all three actions, but the nature of the near-circular orbits guarantees that the stars are already proximate in $J_R$ and $J_z$.} and differences in frequencies $\delta \Omega_\phi$. 
In a static, axisymmetric potential, $\delta J_\phi$ is conserved for all time while $\delta \theta_\phi$ will grow only very slowly, $\simeq \delta \Omega_{\phi} t$. 
These stars will experience similar histories and be sensitive to the same perturbations. \citep[See][for a more complete discussion]{Hunt:2020}. 

With these considerations in mind, we bin our particles in  $(\theta_\phi,J_\phi)$.
Indeed, this has been shown to create cleaner phase spirals both in the real data and in simulations \citep{Li:2021, Hunt:2022}.
Our bins have width $100 \kpckms$ in $J_\phi$ (corresponding to $\simeq 0.4$ \kpc radial bins) and $\pi/8$ radians in $\theta_{\phi}$.
In both simulations, we avoid the inner regions of the galaxy and center our innermost action bin at $J_{\phi} = 1000 \kpckms$ ($R_G \simeq 4.5 \kpc$ for the test particle and N-body simulations, respectively).
In the test particle simulation, the outermost bin is centered at $J_{\phi} = 3000 \kpckms$ ($R_G \simeq 14.6 \kpc$), while in the N-body simulation, the outermost bin is centered at $J_{\phi} = 4000 \kpckms$ ($R_G \simeq 17 \kpc$).
We note for clarity that because the potentials are different in the two simulations, the scaling from $J_\phi$ to $R_G$ also differs.
We perform our analysis using $J_\phi$ bins but show the corresponding $R_G$ where possible in our figures for better intuition.

In total, this means there are 496 disk regions in the N-body simulation and 336 in the test particle one, with 16 bins at each $J_{\phi}$ radius.
We adopt different $J_\phi$ outer limits because of the different disk density profile mentioned in \secref{nbody_sim}.
The N-body simulation is less centrally concentrated meaning it has more stars at higher galactocentric radii (\ie higher $J_\phi$). 
This in turn leads to having enough stars for well-defined phase spirals for $3000 \kpckms < J_\phi < 4000 \kpckms$.

\subsection{Characterizing the phase spiral}
\label{sec:snails}

To characterize the phase spirals, we perform the same analysis for each bin: finding a clean representation of the data using Basis Functions Expansions (BFE's, section \ref{sec:bfes}), and deriving winding times for each bin from these representations (section \ref{sec:winding_time}).

\subsubsection{Quantifying phase spiral properties with basis function expansions} \label{sec:bfes}

\begin{figure*}[t!]
    \centering
    \includegraphics[width=\linewidth]{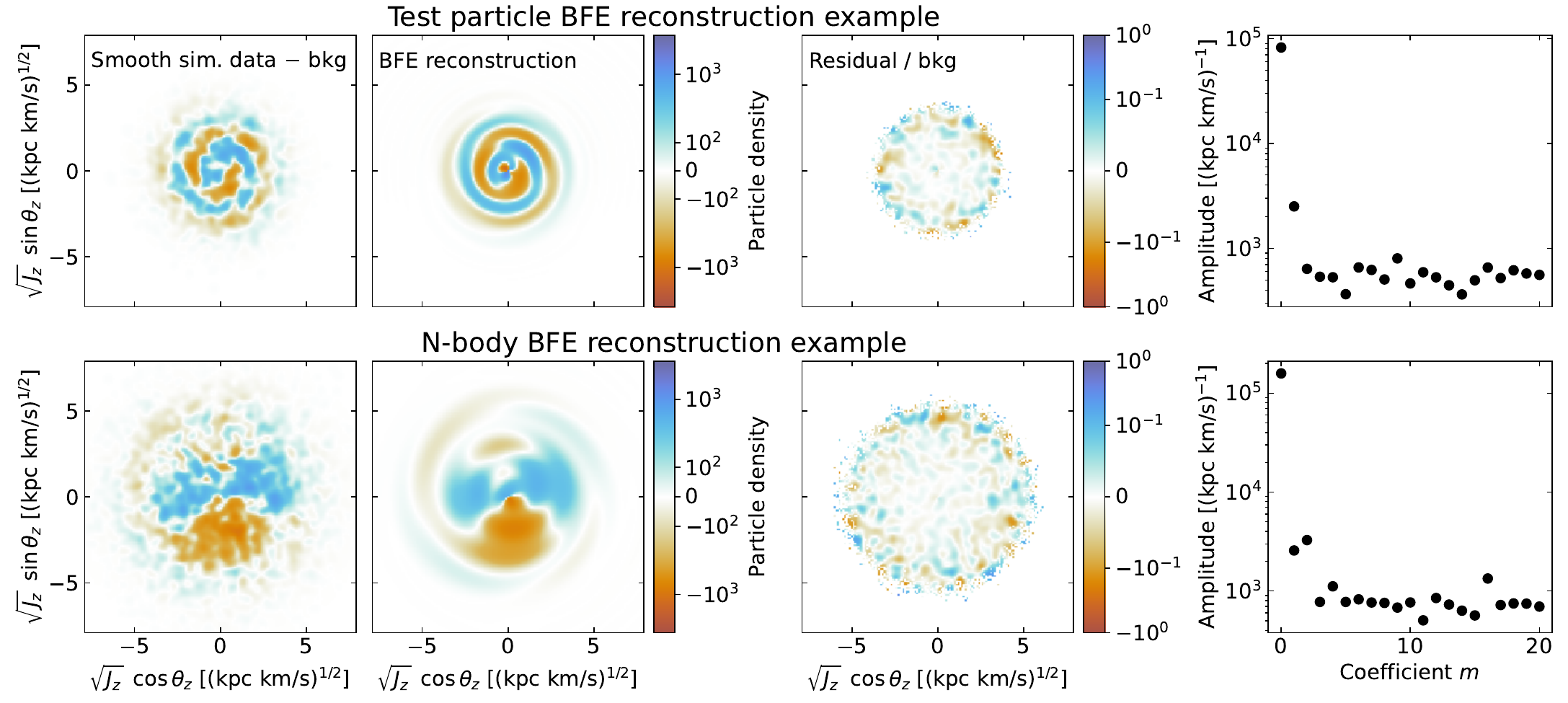}
    \caption{
    Two examples of the BFE reconstructions for phase spirals in the test particle (top row) and N-body (bottom row) simulations.
    In the test particle example, we show the BFE reconstruction for a snapshot taken 0.5 \Gyr 
    after the interaction in a region with center $(\theta_{\phi}, J_{\phi}) = (\pi/2 \textrm{ rad}, 2000 \kpckms)$.
    In the N-body example, we show the BFE reconstruction for a snapshot taken $\simeq 1$ \Gyr after the interaction in a region with center $(\theta_{\phi}, J_{\phi}) = (\pi/2 \textrm{ rad}, 2000 \kpckms)$.
    In each row, the left panel shows the smoothed (using a 2D gaussian filter with $\sigma=0.2 \: (\kpc \km/\second)^{1/2}$) background-subtracted data from the simulation. The second panel shows the 2D reconstruction from the BFE using the $m=1$ and $m=2$ coefficients. The third panel shows the residual when subtracting the first two panels, divided by the background. For clarity, we only perform the residual calculation for pixels containing more than 5 particles.
    The rightmost panel shows the combined amplitude of coefficients for each $m$, defined by $\sqrt{\sum_n A_{nm}^2}$ for $n=0,\ldots,19$.
    The figure demonstrates that the BFEs effectively recover the distribution of particles for different phase space morphologies and separate noise from the signal.
  }
\label{fig:bfe_reconstruction}
\end{figure*}

After projecting particles to ($\theta_z,J_z$), we next perform a Basis Function Expansion (BFE) of their phase-space distribution, $f(\theta_z,J_z)$.
BFEs are a way of capturing the nature of complex fields by expressing them as a linear sum of functions -- the basis set.
In our case, the coefficients of the derived BFEs summarize the information from the particles in each bin using far fewer numbers.

We choose to expand 
\begin{equation} \label{eqn:fourier_laguerre}
    f(\theta_z, J_z) = \sum_{n,m} A_{nm} G_n(J_z) e^{im\theta_z}
\end{equation}
$A_{nm}$ are the BFE coefficients, defined by
\begin{equation}
    A_{nm} = \Big[\frac{\pi}{2}(1+ \delta(m))\Big]^{-1/2} \sum_k G_n(J_{z,k})
    e^{im\theta_{z,k}}
\end{equation}
where $\delta$ is the Dirac-delta function and ($\theta_{z,k}, J_{z,k}$) are the ($\theta_z,J_z$) values for the $k$th particle.


We adopt generalized Laguerre polynomials as our radial basis $G_n$ because the lowest order function $G_0$ is a close match to the equilibrium $J_z$ distribution:
\begin{equation} \label{eqn:gen_laguerre}
    G_n(J_z) = \frac{1}{a\sqrt{n+1}} \exp{\left( -\frac{J_z}{a} \right)}L_n^1 \left( \frac{2J_z}{a} \right).
\end{equation}
$L_n^1$ is the associated Laguerre polynomial of order 1 and degree $n$ while $a$ is the scale length of the vertical action profile of the disk.
In order to reconstruct the distributions with the fewest functions possible, we fit for $a$ in each bin such that the lowest order Laguerre term (\ie the $n=0$ term) is the best fit to the $J_z$ distribution. 
\citet{Weinberg&Petersen:2021} and \citet{Johnson:2023} successfully used this basis to replicate disk structure and we refer the interested reader to those papers for more information about both BFEs in general, and the Fourier-Laguerre basis specifically.

A major advantage of our choice of basis is that the signals we are interested in (1-armed or 2-armed phase spirals) are clearly separated and confined to the $A_{n1}$  and $A_{n2}$ expansion coefficients. This makes the process of identifying one-armed and two-armed features automatic: we do not have to rely on visual inspection or approximate background subtraction.

Using this basis, we can describe each phase spiral with a coefficient series as well as reconstruct the distribution using those coefficients.
We demonstrate the success of these reconstructions by showing one example from the each simulation in \figref{bfe_reconstruction}.
The left panels show a smoothed (using a 2D gaussian filter with $\sigma=0.2 \: (\kpc \km/\second)^{1/2}$) background-subtracted 2-D histogram of stars in the chosen $(\theta_\phi,J_{\phi})$ bin, which is a standard approach to identifying spiral features.
To subtract out the background distribution, we binned the stars by $\sqrt{J_z}$ (100 bins of equal width from $\sqrt{J_z}=0$ to $10$ $\kpc^{1/2} \: \km^{1/2} \:  \mathrm{s}^{-1/2}$) and then subtracted the mean of each bin from the 2D histogram. 
The middle-left panels show the reconstructed distributions from the BFE.
For these reconstructions, we limit the number of Laguerre functions to 20 ($n=0,\ldots,19$).
We also use only the $m=1$ and $m=2$ Fourier modes to isolate the one-armed and two-armed features in each region.
The middle-right panels are the residuals of the first two columns divided by the background.
In this case the background is defined as the BFE reconstruction from the $m=0$ coefficients.
We can clearly see that these are noise-dominated.
In the right panels, we show the combined amplitude of the coefficients for each $m$, defined by $\sqrt{\sum_n A_{nm}^2}$ 
for $n=0,\ldots,19$, demonstrating that $m>2$ is dominated by noise.
In addition to these two examples, we visually verify that this BFE reconstruction technique is robust for many different $(\theta_z,J_z)$ morphologies and proceed to make the calculation for every region at every timestep.

With the quantitative description of phase spirals that BFE reconstructions provide us, we can derive a variety of phase spiral properties. 
For example, the basis function coefficients directly give us amplitudes of one-armed and two-armed spirals.
In a companion paper (Tavangar et al. in prep),  we will explore the correlations between phase spiral amplitudes across the disk during the simulation.
Additionally, using the reconstructions, we can derive pitch angles and (equivalently) winding times, which will be the focus of this work.

\subsubsection{Deriving winding times} \label{sec:winding_time}

One descriptor of the morphology of a spiral is the tightness of the winding, which can be quantified by its pitch angle, $\psi$. 
This can be translated to a quantity of relevance for dynamics and history by calculating the winding time \citep[which we will call $T_{\rm fit}$ following the notation in][]{Asano&Antoja:2025}, or the time it would take for the spiral to attain this pitch angle, starting from an asymmetric distribution and evolving in the absence of self gravity through phase-mixing alone.

Our primary goal in this paper is to understand the effectiveness of using $T_{\rm fit}$ to recover the time at which the perturbation which incited a spiral actually occurred.
In this section and for the remainder of this work, we will focus on the one-armed phase space spirals.
This is both because they are the dominant spirals in our simulations and because \citet{Widrow:2023} showed that their evolution in the presence of self-gravity is simpler than that of the two-armed spirals, which we leave to future work.

Our method for deriving $T_{\rm fit}$ for a given spiral is very similar to that adopted by previous works \citep[\eg][]{Darragh-Ford:2023, Asano&Antoja:2025}, with the key difference being that we are working directly using the $m=1$ terms in our BFE rather than fitting functions to (noisy) particle data.
The method relies on the fact that each phase spiral has a ridgeline of highest density in ($\theta_z,J_z$) space. 
For a one-armed spiral, each $J_z$ value has a single $\theta_z$ where the phase spiral density is highest, which we call $\theta_{z, \textrm{max}}$.
Furthermore, there is a straightforward relationship between the phase difference on the ridgeline ($\Delta\theta_{z, \textrm{max}}$) for two $J_z$ values, say $J_{z1}$ and $J_{z2}$, the difference in the vertical frequency ($\Delta{\Omega_z}$) at those $J_z$ values, and $T_{\rm fit}$:
\begin{equation} \label{eq:wind_time}
T_{\rm fit} = \frac{\theta_{z, \textrm{max}}(J_{z1}) - \theta_{z, \textrm{max}}(J_{z2})}{\Omega_{z}(J_{z1}) - \Omega_z(J_{z2})} \equiv \frac{\Delta\theta_{z, \textrm{max}}}{\Delta\Omega_z}.
\end{equation}

In our work, we perform this $T_{\rm fit}$ calculation by finding $\Omega_z$ and $\theta_{z, \textrm{max}}$ as a function of $J_z$ for each phase spiral.
To do this, we first split the BFE reconstruction from $m=1$ into 96 $\theta_z$ bins and 100 $\sqrt{J_z}$ bins from 0 to 10 $\kpc^{1/2} \: \km^{1/2} \:  \mathrm{s}^{-1/2}$.
The $\Omega_z$ calculation is straightforward: we take the median $\Omega_z$ of the particles in each $\sqrt{J_z}$ bin.
Getting $\theta_{z, \textrm{max}}$ as a function of $J_z$ is more involved.
For each $\sqrt{J_z}$ bin, we first find the $\theta_{z}$ bin with the highest density 
This creates a ridgeline in $(\theta_z,\sqrt{J_z})$ space, to which we fit a logarithmic spiral:
\begin{equation}
    \theta_{z,{\rm max}} = \frac{\ln{\sqrt{J_z}}}{\tan{\psi}} + \theta_0
\end{equation}
where $\theta_0$ is the spiral's phase angle.
This choice of functional form is arbitrary but from visual inspection of many phase spirals in our simulations, the logarithmic spiral is an excellent match, especially when discounting $\sqrt{J_z} < \sqrt{a}$, which we do here.

The fitted spiral allows us to use Equation~\ref{eq:wind_time} because we can now calculate $\theta_{z, \textrm{max}}$ for any two choices of $J_z$.
For the lower (\ie inner) $J_z$ value, we select $a$.
For the upper (\ie outer) $J_z$ value, we select the highest $J_z$ for which the corresponding $\sqrt{J_z}$ bin contains at least 96 particles.
This latter choice is somewhat arbitrary but ensures we only include regions with a meaningful number of stars.

\section{Results} \label{sec:results}

\subsection{Test particle phase spiral winding} \label{sec:tp_results}

\begin{figure*}[t!]
    \includegraphics[width=1.15\linewidth]{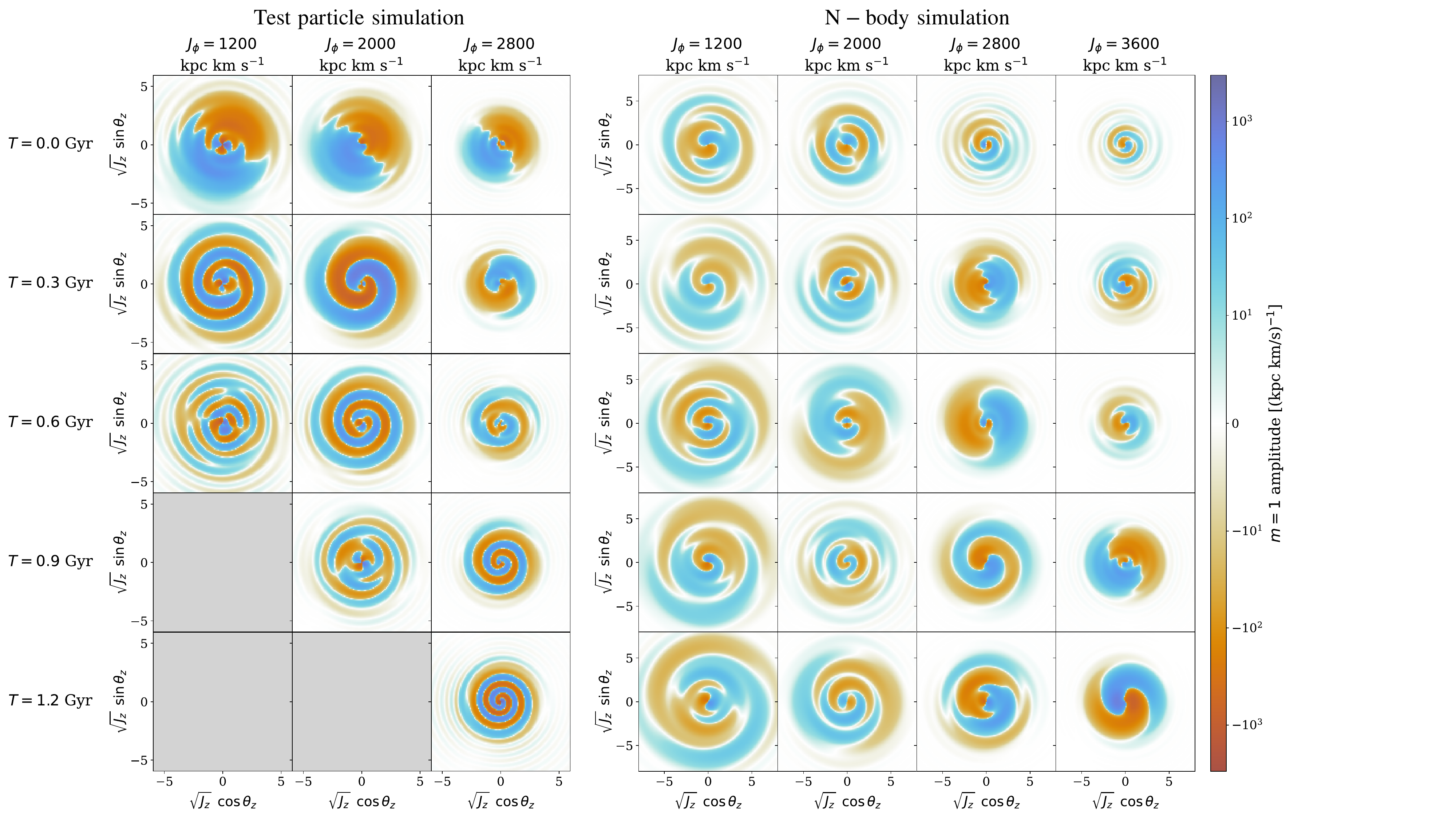}
    \caption{{\it Left:} BFE reconstructions ($m=1$ only) of the phase spiral in the test particle simulation. Each row shows phase spirals at different simulation times, where $T=0$ \Gyr is the disk crossing time. Each column shows a a different region, with $J_\phi$ increasing from left to right. For each $J_\phi$ we choose the bin with azimuthal center $\theta_\phi=0$. At the same timestep, We see more wound up phase spirals in the inner disk. We gray out the panels where the spiral is too wound up for the resolution of the simulation, leading to an nonphysical reconstruction.
    {\it Right:} The same as the left grid but for the N-body simulation. Here we add an additional column because we have sufficient particles to reconstruct phase spirals out to a larger $J_\phi$. These phase spirals are noisier due to the more complex simulation and interaction. They are also noticeably less wound up than in the test particles case, especially in the inner disk. 
    }
\label{fig:tp_nbody_grid}
\end{figure*}

We first examine the phase spirals in the test particle simulation as a sanity check to ensure our $T_{\rm fit}$ calculation from the BFE reconstructions gives reasonable results.
As described in \secref{test_sim} these phase spirals form following the disk crossing of a satellite $\simeq 15$ kpc from the galaxy's center.
In the left grid of \figref{tp_nbody_grid}, we show the resulting one-armed phase spiral reconstruction for 15 example spirals at three different $J_\phi$ and five different times.
As expected, we find that the phase spirals wind up over time until they become so wound that their structures can no longer be resolved.
This is the natural process of phase mixing as the system returns to a quasi-equilibrium state.

\begin{figure*}[t!]
    \centering
    \includegraphics[width=0.85\linewidth]{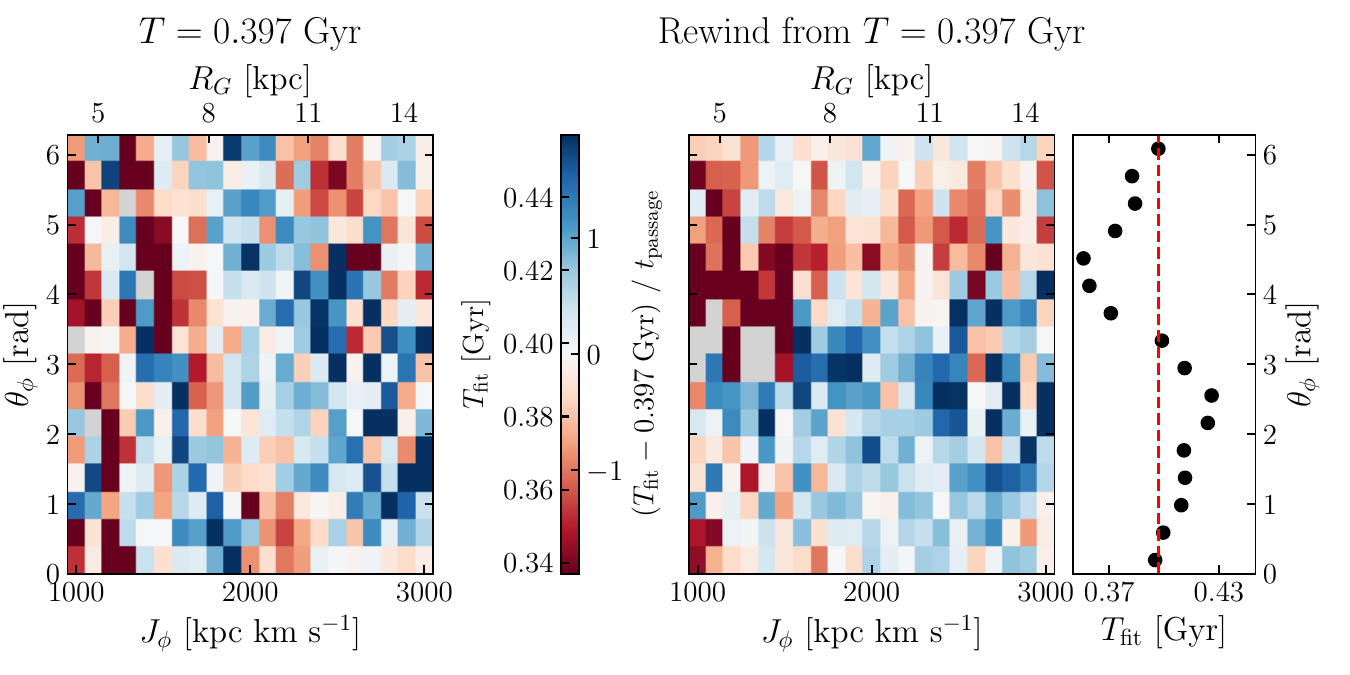}
    \caption{{\it Left:} $T_{\rm fit}$ calculated for each bin 0.4 \Gyr after the disk crossing in the test particle simulation. 
    {\it Middle:} The same as the left panel except that the regions have been rewound to where they were at the time of the interaction. The diagonal ridges of red or blue in the left panel (which would appear as a ``macro-spiral'' in a face-on view of the disk) unwind into a dipole. This occurs because of differences in when each region experiences the highest amplitude perturbation from the satellite, as explained in \secref{tp_results} and \citet{Gandhi:2022}. We note that the perturber crosses the disk at $\theta_\phi = 0$ (or $\pi$), exactly where the dipole split appears.
    {\it Right:} The mean $T_{\rm fit}$ values for each $\theta_\phi$ row in the rewound (middle) plot. The true time since the disk crossing ($0.397 \Gyr$) is denoted with the red dashed line.}
\label{fig:tp_dipole}
\end{figure*}

Using the method described in \secref{winding_time}, we derive the perturbation time from each spiral in the disk.
We show the results in \figref{tp_dipole}, which visualizes them in two ways.
In the left panel, we show $T_\mathrm{fit}$ for the spiral in each region approximately 0.4 \Gyr after the satellite disk crossing. 
The colorbar shows the winding time, with $T_\mathrm{fit}$ values indicating perturbation times slightly before (after) the disk crossing shown in blue (red).
We indicate this winding time both in \Gyr and in units of $t_{\rm passage} \equiv r_{\rm sat}/v_{z,\rm sat}$ with both $r_{\rm sat}$ and $v_{z,{\rm sat}}$ taken at the disk crossing time.
We derive the expected $T_\mathrm{fit}$ value of $\simeq 0.4$ \Gyr to within $50 \Myr$ for nearly every region in the disk.
However, we notice that while the perturbation times are constrained to a narrow range, they appear to be bimodal, with very few of the regions having $T_\mathrm{fit}$ of exactly $0.4 \Gyr$.
Instead, they are mostly slightly earlier or later and regions with similar derived times seem to fall along diagonal ridges.
Given that regions at different $J_\phi$ values have different orbital periods and therefore different orbital frequencies, we can apply the same physics to these diagonal ridges as for the phase spiral winding time calculations.
In other words, we can rewind the disk spiral in the same way we did the phase spiral.
We do this to create the second and third panels of \figref{tp_dipole},
where we see a bimodality much more clearly and observe that the resulting structure is a dipole along $\theta_\phi = 0,\pi$.
The fact that this dipole appears when rewinding to the disk crossing time shows a separate way -- along with rewinding individual phase spirals -- to prove that the satellite passage is responsible for the phase spirals.
We will build on this idea of rewinding large-scale features in our companion paper exploring phase spiral amplitudes in simulations.

To explain why this dipole appears, we repeat the discussion outlined in \citet{Gandhi:2022}.
Regions which are moving towards the impact point of the disk crossing experience a larger force after the satellite has crossed the disk because they are that is when they are closer to it.
On the other hand, regions moving away from the impact point experience a larger force before the disk crossing. 
This leads to a slight leading/trailing offset in the times at which regions of the disk ahead/behind the satellite's disk crossing experience the interaction.
In our test particle simulation, stars rotate counter-clockwise and the satellite goes from negative to positive $z$ values.
The dipole we see matches this physical picture.

\begin{figure*}[t!]
    \centering
    \includegraphics[width=\linewidth]{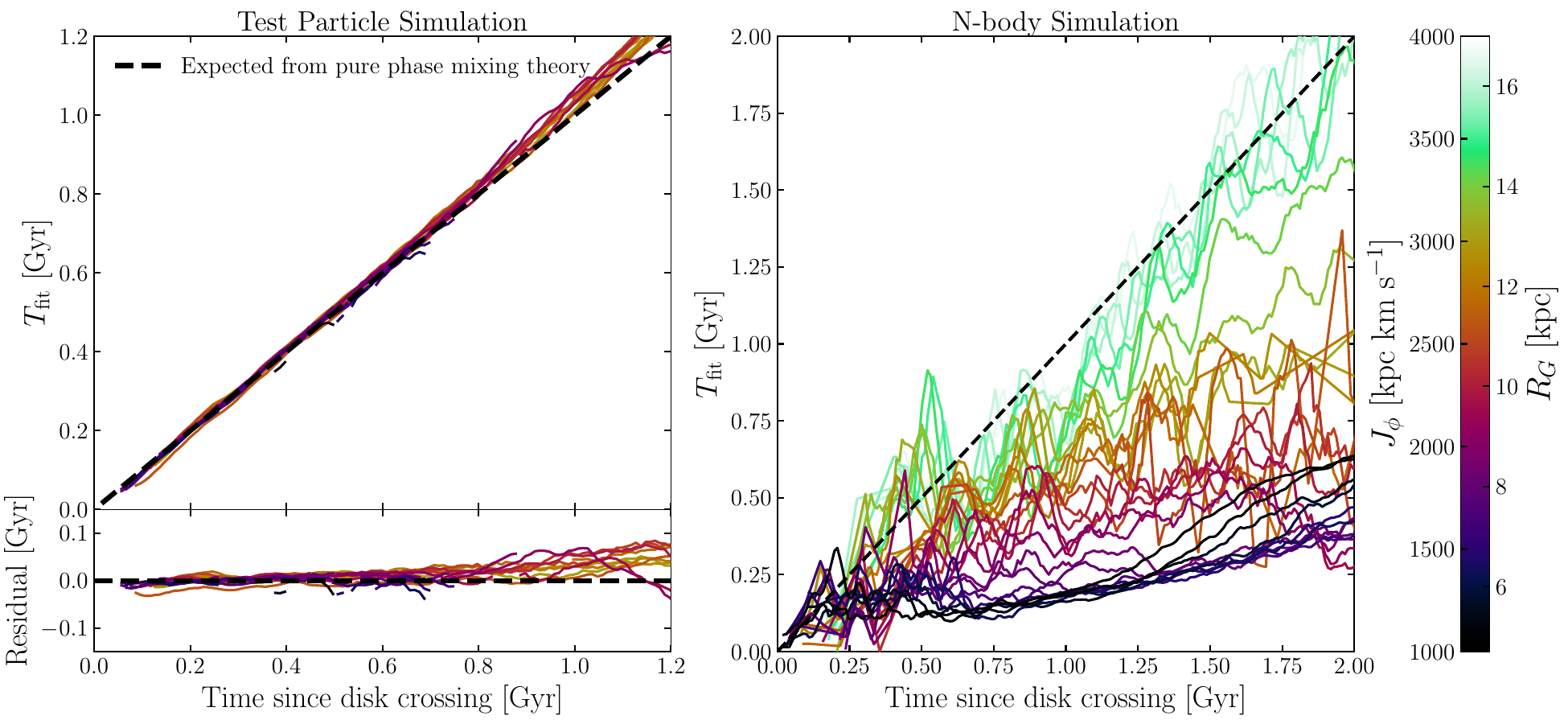}
    \caption{{\it Top left:} The $50 \Myr$ moving average of the median $T_{\rm fit}$ value for each $J_\phi$ at each test particle simulation timestep. The dashed black line corresponds to the expected $T_{\rm fit}$ at each timestep from pure phase mixing theory. As discussed in \secref{tp_results}, the phase spirals in this simulation eventually wind up too much for our resolution to capture them, at which point the $T_{\rm fit}$ calculations are meaningless and fail. This is the reason why some of the curves are not continuous or do not span the entire x-axis.
    \textit{Bottom left:} The residual of the top left plot with respect to the expected $T_{\rm fit}$ from pure phase mixing theory. This can also be interpreted as the difference between the derived perturbation time from $T_{\rm fit}$ and the true disk crossing time.
    {\it Right:} The $50 \Myr$ moving average of the median $T_{\rm fit}$ value for each $J_\phi$ at each N-body simulation timestep. In our median calculations, we only include regions with $T_{\rm fit}$ values between 0 and 2 times the value expected from phase mixing theory. We see clearly here that while there are winding delays at all $J_\phi$, the phase spirals in the inner disk are far more affected by n-body interactions. 
    Note: the conversion from $J_\phi$ to $R_G$ in the colorbar is done according to the N-body simulation. The test particle simulation does not have exactly the same scaling (see the two x-axes in \figref{tp_dipole} for a visualization of the test-particle scaling).
    }
\label{fig:tp_nbody_winding}
\end{figure*}

Finally, in the left panels of \figref{tp_nbody_winding}, we plot the $50 \Myr$ moving average of the median $T_{\rm fit}$ in each $J_\phi$ annulus at each timestep of the simulation, along with the residual relative to the expected value.
As anticipated for this simulation, we find that the pure phase mixing theory describes the post perturbation state of one-armed phase spirals throughout the disk very well\footnote{The discontinuities at later times for the inner disk occur because the one-armed phase spiral eventually phase mixes completely and is no longer distinguishable from the equilibrium background, causing the $T_{\rm fit}$ calculation to fail.}.

\subsection{N-body phase spiral winding} \label{sec:nbody_results}

\begin{figure*}
    \centering
    \includegraphics[width=\linewidth]{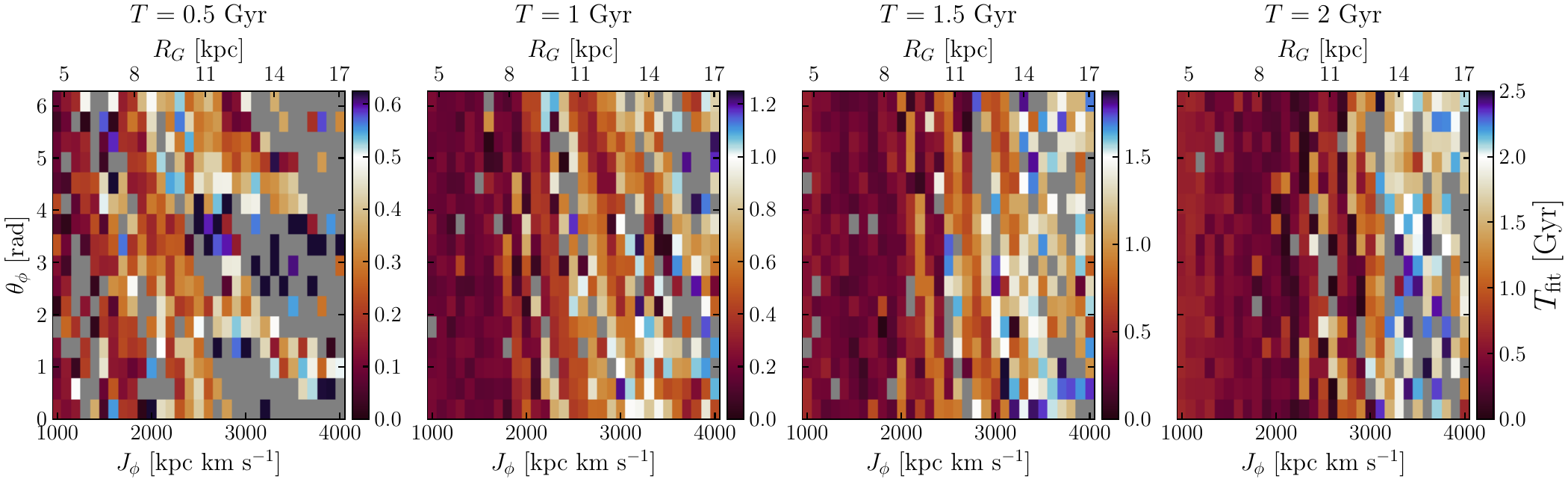}
    \caption{Four panels showing $T_\mathrm{fit}$ for every region of the disk at different simulation times. Red values indicate there is a delay in phase spiral winding. We see significant delays at all timesteps, but also large variation with $J_\phi$. Regions where the $T_\mathrm{fit}$ calculation failed are shown in gray. Regions in white recover the correct winding time.}
    \label{fig:nbody_tfit}
\end{figure*}

We now analyze how phase spiral winding differs in a self-consistent simulation.
Since we are trying to isolate this effect, we focus on the section of the N-body simulation most similar to the one analyzed in the test particle case: the period between the first and second passages of the satellite through the disk.
This allows us to examine only the impact of a single satellite disk crossing, as we did in the test particle case.
Despite this limitation, we acknowledge that there are still differences in the orbits of the satellites in the two simulations which could impact the phase spiral.
First, the disk crossing location is about twice is far from the galactic center in the N-body case ($\simeq 15$ \kpc \vs $\simeq 34$ \kpc).
Second, since the satellite remains bound, it is slower during the flyby.
Both of these differences contribute to the perturbation being less impulsive in the N-body case.

With these differences in mind, we perform a similar analysis to \secref{tp_results}.
In the right grid in \figref{tp_nbody_grid}, we show 20 example phase spirals at different $J_\phi$ values and times.
We add a column relative to the analogous grid in the test particle analysis because we have higher particle resolution in the outer disk of our N-body simulation, allowing us to extend our $J_\phi$ bins to $J_\phi = 4000 \: \kpckms$ ($R_G \simeq 17 \kpc$).
The clearest takeaway from these phase spirals is that they are not as wound up as their counterparts in the test particle grid.
This is especially apparent in the inner disk where we still see a clear spiral arm $\simeq 0.8 \Gyr$ after the perturbation whereas the analogous region had completely phase mixed by that point in the test particle simulation.

To confirm that these less wound spirals are evidence of a delay in winding rather than differences in the potentials, we plot $T_\mathrm{fit}$ for all the disk regions at four different timesteps in \figref{nbody_tfit}.
Here we see that spirals throughout the disk are less wound up than they would be under test particle conditions, with $T_\mathrm{fit}$ consistently lower than the elapsed time since the perturbation.
We also see that $T_\mathrm{fit}$ depends strongly on $J_\phi$, with the inner disk generally having lower $T_{\rm fit}$ values (\ie experiencing more delay) than the outer disk.

We show this a different way in the right panel of \figref{tp_nbody_winding}, where we plot the $50 \Myr$ moving average of the median derived winding times\footnote{To avoid including regions where the $T_\mathrm{fit}$ calculation clearly failed, we only consider regions with $T_\mathrm{fit}$ values between 0 and 2 times the value expected from phase mixing theory.} for the regions in each $J_\phi$ annulus.
Our choice to smooth over $50 \Myr$ is to help visualize the trends in our results.
In contrast to the test particle simulation, $T_{\rm fit}$ mostly does not increase as expected from phase mixing theory and varies significantly with $J_\phi$.
In the inner disk ($J_\phi \lesssim 2000 \kpckms$), we observe significant delays in phase spiral winding, with the largest delays in the innermost regions.
At slightly higher $J_\phi$ ($2000 \kpckms \lesssim J_\phi \lesssim 3000 \kpckms$), we generally see higher $T_{\rm fit}$ values, meaning less winding delay.
Meanwhile, in the outer disk ($3000 \kpckms \lesssim J_\phi \lesssim 4000 \kpckms$) the derived $T_{\rm fit}$ does give approximately the correct time since perturbation, meaning there are minimal winding delays.
This panel has two additional features of interest: (i) varying average slopes for $T_{\rm fit}$ as a function of $J_\phi$, and (ii) periodic oscillations in $T_{\rm fit}$ for $J_\phi \gtrsim 2000$.
We explore both of these, along with the delays at the start of winding, by building toy models in \secref{explaining_delay}.

\section{Towards Understanding the features in the N-body winding evolution} \label{sec:explaining_delay}

Prior work on the evolution of phase spirals has shown that accounting for self-gravity -- either by using self-consistent simulations or mimicking N-body interactions some other way -- leads to winding delays \citep{Darling:19a, Darling:19b, Darling:21, Bennett&Bovy:2021, Widrow:2023, Bland-Hawthorn_Tepper-Garcia:2021, Darling:24, Asano&Antoja:2025}.
The results in this paper more explicitly demonstrate that phase-spirals in N-body simulations exhibit (i) winding delays, (ii) winding slowdowns, and (iii) oscillations in the winding rates, as shown in $T_{\rm fit}$ curves in the right panel of \figref{tp_nbody_winding}. All of these effects vary as a function of $J_\phi$.

In this section, we explore just one aspect of the dynamics influencing phase-spiral evolution --- namely, how an evolving galactic disk can affect the winding of a group of particles, traveling together in one of our ($\theta_\phi,J_\phi$)-bins.
In \secref{local_global}, we first examine the $J_\phi$ dependence of the nature of winding by considering local and global contributions to particle accelerations in different regions of the disk; 
in \secref{toys}, we go on to use toy models to build intuition for how the evolving disk potential might delay, slow-down or cause oscillations in the winding rates and in 
\secref{assess} we assess how much of what we have observed in our models can be attributed to the evolving disk alone.

\subsection{Local versus total accelerations to explain the $T_{\rm fit}$ dependence on $J_\phi$} \label{sec:local_global}

\begin{figure*}[t!]
    \centering
    \includegraphics[width=\linewidth]{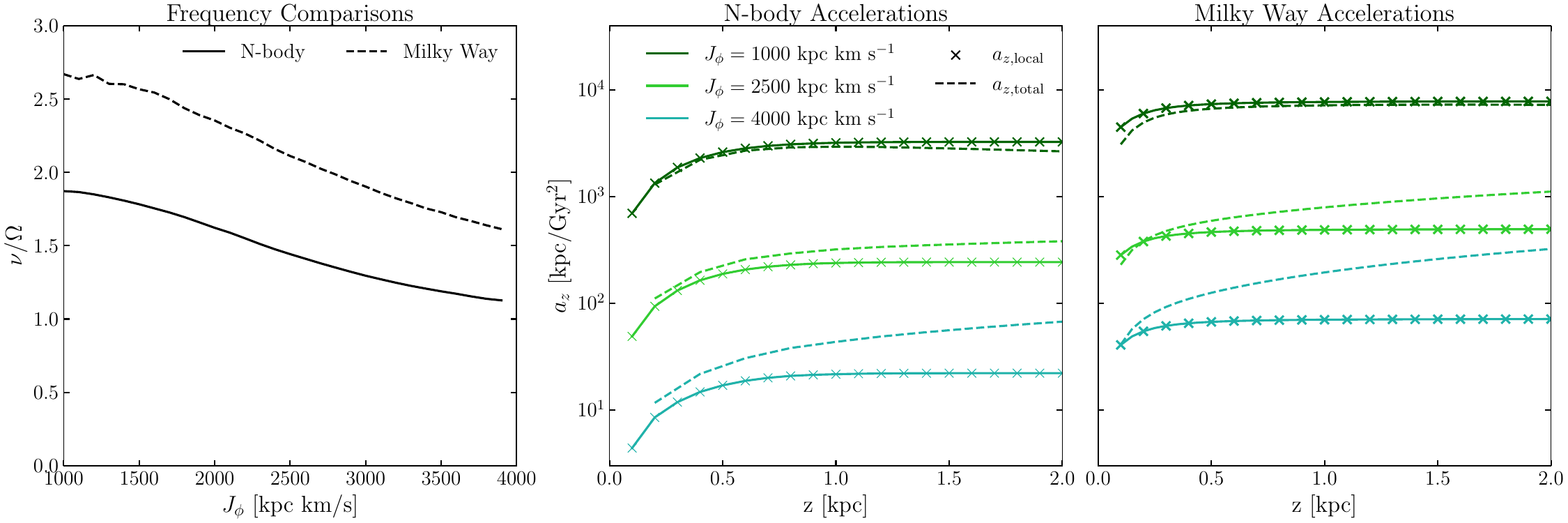}
    \caption{\textit{Left:}  The ratio of vertical and azimuthal frequencies for stars on circular orbits as a function of $J_\phi$ for the N-body simulation (solid line) and the MW (dashed line).
    When $\nu/\Omega \equiv 1$, the potential is spherical.
    \textit{Middle:} The local component of the vertical acceleration (solid lines with crosses) compared to the total vertical acceleration (dashed lines) as a function of vertical height for three different regions of the disk in the N-body simulation. The innermost region ($J_\phi = 1000 \kpckms$) is plotted in sea green, the middle region ($J_\phi = 2500 \kpckms$) in lime green, and the outer region ($J_\phi = 4000 \kpckms$) in dark green.
    We note that in the inner disk, our estimate for the local acceleration actually exceeds the full disk acceleration as our assumption of an infinitely extended, high-surface density disk becomes unrealistic (see \secref{local_global}).
    \textit{Right:} The same as the middle panel but for the MW.}
\label{fig:freq_and_accel}
\end{figure*}

The ratio $\nu/\Omega$ (where $\nu = \Omega_z(z=0)$ and $\Omega = \Omega_\phi(z=0)$) provides a useful assessment of the dominance of the disk component in a galactic potential. It is identically unity in perfectly spherical systems, and increases with the oblateness of the system. 
The left panel of \figref{freq_and_accel} shows $\nu/\Omega$ for circular orbits in both our N-body model and the MW.
The fact that this ratio is highest towards the center of the disk and decreases to almost unity in the outskirts is already an indication of the decreasing importance of disk potential with increasing $J_\phi$.

We can assess the situation more carefully by comparing the vertical acceleration due to all the disk particles in our simulation (the ``total'' acceleration, $a_{\rm total}$) to the acceleration from nearby stars alone (the ``local'' acceleration, $a_{\rm local}$).
The total acceleration is calculated directly from all the particles in the simulation using the \texttt{pytreegrav} package \citep{pytreegrav} for determining gravitational potentials and accelerations of N-body simulations.
The scale of the local acceleration perpendicular to the $z=0$ plane is estimated by $2\pi G\Sigma_0$, appealing to the acceleration due to an infinitely thin disk where $\Sigma_0$ is the surface density.
We calculate the surface density of stars enclosed within $\pm z$ of the plane for an annulus at radius $R$ and width $\Delta R$ and adopt:
\begin{equation}
  a_{z, \rm local} (R,z) =  \frac{G}{R \Delta R} \sum_{i, |z_i|<z} m_i
\end{equation}

The results are shown in the middle panel of \figref{freq_and_accel}.
Total accelerations as a function of $z$ are given by dashed lines while the local accelerations are shown by solid lines with crosses. The different colors correspond to different parts of the disk.
In this panel it is clear that the local contribution to the acceleration is less significant in the outer disk than in the middle or inner disk (visualized as a larger gap between the crosses and the dashed line).
Hence, even if a local patch of the outer disk is perturbed away from the global midplane, the acceleration field changes little and the phase spiral in this patch will form similarly to a test particle simulation.
In other words, the phase spiral will wind up as in a test particle simulation -- with no delay -- which matches our results.

In contrast, local patches in the middle and inner disk account for nearly all the vertical acceleration.
Indeed, our estimate for the local acceleration actually exceeds the full disk acceleration in the inner disk as our assumption of an infinitely extended, high-surface density disk becomes unrealistic.
Hence, in the inner disk we expect the acceleration field to be significantly perturbed if the disk itself is perturbed. 
Consistent with this, in our N-body results we find that the phase spirals in these regions have large winding delays. 

We explore one physical mechanism that might cause this delay in \secref{toy1}.

\subsection{Exploring individual $T_{\rm fit}$ curve features with toy models} \label{sec:toys}

Our basic setup for these toy models is a 1D system of $10^5$ test particles in a specified potential, meant to represent a single $(\theta_\phi,J_\phi)$ bin.
These particles are initialized in an $\ln \cosh$ potential (the same as the vertical potential of the N-body simulation) with scale height $z_h$ and are initially in equilibrium around the $z=0$ midplane.
In each variation of this toy model, we impart a velocity kick on these particles equal to the velocity dispersion $\sigma_z$.
We then allow the particles to evolve but change the potential in prescribed ways corresponding to different physical mechanisms in order to recreate the various features in \figref{tp_nbody_winding}. Specifically, we explore effects of the disk midplane reforming as it relaxes from a perturbation (\secref{toy1}); its vertical structure evolving, e.g. due to dynamical heating (\secref{toy2}); and oscillations, e.g. due to breathing modes (\secref{toy3}).

\subsubsection{Toy model \#1: creating a delay at the start of winding} \label{sec:toy1}

\begin{figure}[t!]
    \centering
    \includegraphics[width=\linewidth]{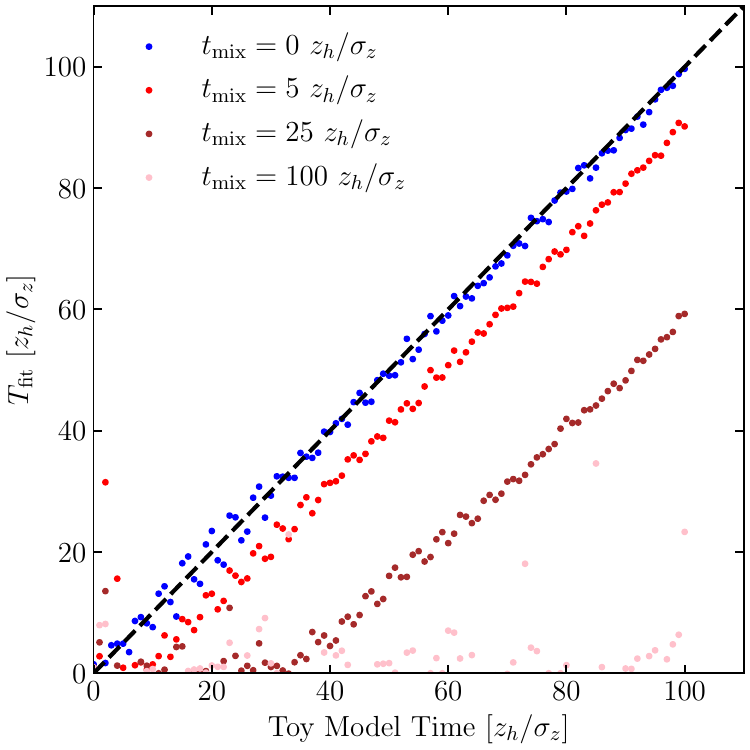}
    \caption{The results of running our first 1-D dimensionless toy model (see \secref{toy1}) with different $t_\mathrm{mix}$ values, which specifies the characteristic timescale on which the original disk midplane repopulates itself due to phase-mixing and relaxation. Both time axes are in dimensionless units of the scale height over the velocity dispersion. We clearly see that this quantity has a significant effect on the delay in phase spiral formation.}
\label{fig:toy}
\end{figure}

In \secref{local_global} we showed a correlation between regions where local patches dominate the acceleration field and regions where there is a prominent delay in the phase spiral winding.
We now set out to demonstrate that there is a link between the two by developing an intuitive picture for one physical mechanism that might cause a delay.
Consider the experience of a set of particles in one of our $(\theta_\phi,J_\phi)$ bins that has been maximally perturbed from the midplane during an interaction compared to other bins of the same $J_\phi$.
Initially, all the particles in the same spatial patch as those in the $(\theta_\phi,J_\phi)$-bin are perturbed in the same direction and move together, with their coherence reinforced by their N-body interactions. 
However, disks are not solid bodies, so over enough time we expect phase-mixing and relaxation to gradually change the particles that are in the same patch as the $(\theta_\phi,J_\phi)$-bin we are following. 
In particular, an increasing percentage of particles that are spatially coincident with the bin will not have experienced the same initial perturbation and will on average be centered on the original disk plane.
In other words, we expect the disk plane to re-establish itself and the importance of local accelerations for our original perturbed stars to decrease.

We can mimic this expectation using a two-component potential
\begin{equation}
  \begin{aligned}
    \Phi \:=\: &A_0 e^{-t/t_{\textrm{mix}}} \ln \cosh (\frac{z-z_{\rm local}}{z_h}) \: + \\
    &A_0(1-e^{-t/t_{\textrm{mix}}}) \ln \cosh (\frac{z}{z_h})
  \end{aligned}
  \label{eqn:phikick}
\end{equation}
where $z_{\rm local}$ is the mean position of the test particles, calculated at each timestep, and $t_{\rm mix}$ is the mixing time, which encodes the characteristic timescale of re-population.
We now report the results of simulating the equations of motion of our test particle ensemble subject to the gravitational potential in Equation~\ref{eqn:phikick}.

The model elucidates a few important points, all shown in \figref{toy}.
First, we find that $t_{\rm mix}$ determines the length of the delay in phase spiral formation.
Specifically, the spiral starts forming after $\simeq 1.5$ mixing times.
When $t_{\rm mix}$ is very large, the spiral never winds up at all.
Second, once the phase spiral begins forming, it forms at the expected rate from a test particle simulation (\ie the winding rate is unaffected once it starts).
This simple toy model demonstrates that in regions where the local contribution to the acceleration is dominant, we can expect significant delays in phase spiral winding.

We now roughly estimate $t_{\rm mix}$ for particles in our innermost $J_\phi$ bin ($950 \kpckms <J_\phi<1050 \kpckms$).
We can calculate the mixing time by defining $t_{\rm mix}$ as the amount of time it takes for two stars with a $J_\phi$ difference of $50 \kpckms$ (half a $J_\phi$ bin width) to separate by $\pi/4$ radians (two $\theta_\phi$ bin widths) in azimuthal angle.
We note that the precise chosen numbers are arbitrary, to give some indication of $t_{\rm mix}$.
With this definition, we find $t_{\rm mix}$ in the inner disk to be $\simeq 350$ \Myr.
This number only increases for $t_{\rm mix}$ calculations further out in the disk, where azimuthal separation occurs more slowly.
Based on our toy model, this implies that in the absence of global accelerations, we would expect winding delays of several hundreds of Myrs throughout the disk.

We check our N-body results for evidence of this physical picture and find two pieces of evidence supporting it.
First, the inner disk ($J_\phi \lesssim 2000 \kpckms$) has very little winding for at least $\simeq 1$ \Gyr after the disk crossing, as we would expect for regions dominated by local accelerations with long mixing times.
Second, in the right two panels of \figref{nbody_tfit} we see that $T_\mathrm{fit}$ decreases with increasing $J_\phi$ in the inner regions where local acceleration is dominant.
One can also see this at late times in \figref{tp_nbody_winding} where the innermost regions (blackest curves) have higher $T_{\rm fit}$ values than those slightly further out (purple curves).
This matches the expectations from our toy model because $t_{\rm mix}$ increases with $J_\phi$, leading to longer delays.

\subsubsection{Toy model \#2: creating different $T_{\rm fit}$ slopes} \label{sec:toy2}


For a given region, the $T_{\rm fit}$ calculation (see Equation~\ref{eq:wind_time}) depends on the vertical potential of that region at the time of calculation.
It does not account for the fact that the structure of the disk may be evolving, for example as the disk heats in response to an interaction. 
For example, let us suppose that we have a harmonic vertical potential in which stars evolve for some time after being perturbed.
In such a potential, because all stars oscillate with the same frequency, there will be no winding at all.
If we suddenly change the potential to an $\ln \cosh$ potential, a phase spiral will start to form.
In this case, the calculated $T_{\rm fit}$ would be the time since the potential changed, as opposed to the time since the perturbation.

\begin{figure}[t!]
    \centering
    \includegraphics[width=\linewidth]{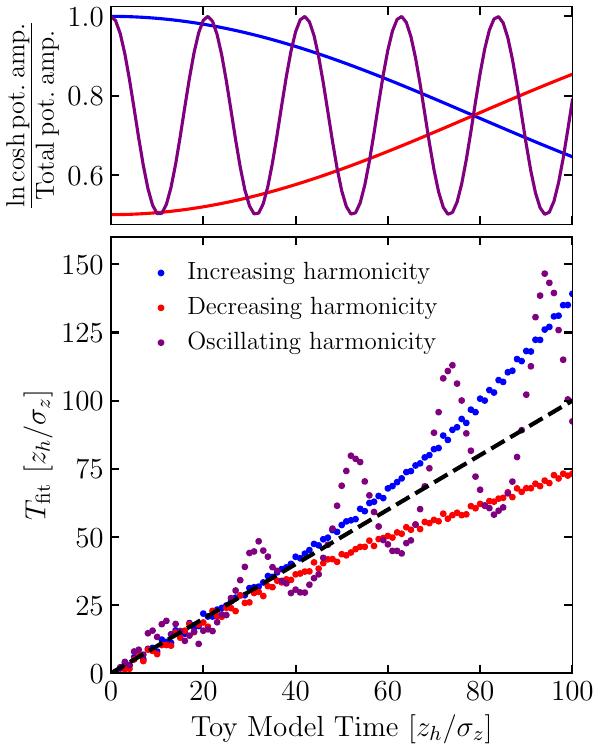}
    \caption{The results of our toy model to explain one way to create the shallower slopes and oscillations in  $T_{\rm fit}$ curves in \figref{tp_nbody_winding}. In these toy model runs, we vary the contributions of two harmonic and anharmonic components to the total potential over time and examine the resulting evolution in $T_{\rm fit}$. \textit{Top:} The evolution over time of the ratio of the anharmonic component of the potential to the total potential (\ie $f$ in Equations~\ref{eq:harmonic_anharmonic} and \ref{eq:s}) for three different runs of our toy model.
    {\it Bottom:} $T_{\rm fit}$ for each of the three runs. }
\label{fig:toy_pot_change}
\end{figure}

To illustrate this, we construct our toy model with a more realistic version of this example.
Instead of the potential changing suddenly from one form to another, we allow it to vary slowly and continuously throughout the simulation.
In detail, we define the potential as: 
\begin{equation} \label{eq:harmonic_anharmonic}
    \Phi = f \times \sigma_z^2 \ln \cosh (\frac{z}{z_h}) + (1-f)\times \frac{1}{2}\nu^2 z^2
\end{equation}
where $\nu=1$ inverse time units and $f$ varies sinusoidally between 0.5 and 1 as a function of time:
\begin{equation} \label{eq:s}
    f(t) = 0.25\times(3 \pm \cos(\omega t))
\end{equation}
The $\pm$ sign in this equation allows us to either increase the harmonic component from 0\% of the potential to 50\% or decrease it from 50\% to 0\%.
We set $\omega=0.02$ inverse time units (period of oscillation $\simeq 300$ time units) to ensure the evolution of $f$ would be monotonic for the duration of the simulation.
The value of $f$ over the course of the model is shown in the top panel of \figref{toy_pot_change} in blue and red, respectively.

The bottom panel of \figref{toy_pot_change} shows the resulting $T_{\rm fit}$ calculation throughout the simulation.
As expected from our thought experiment above, increasing the harmonicity of the potential over time can increase the $T_{\rm fit}$ slope while decreasing it can decrease the $T_{\rm fit}$ slope.
Therefore, one explanation for the average $T_{\rm fit}$ slope in our N-body model being less than 1 for most $J_\phi$ is that the vertical potential in those regions is becoming less harmonic over time.

While this result is found when assuming a specific shape for the potential, it can be stated more generally in terms of frequencies. 
In our case, we find that $T_{\rm fit}$ underestimates the true time since perturbation. 
If this is due to disk evolution, it suggests that the frequency difference across the $z_{\rm max}$ amplitudes of the phase spiral has increased over time. 
This picture is at odds with the physical intuition that the net effect of a perturbation will be to heat the disk, leading to the vertical frequency gradient decreasing. 
Clearly, this topic merits further investigation.

\subsubsection{Toy model \#3: creating oscillations in $T_{\rm fit}$ over time} \label{sec:toy3}

Finally, creating oscillations in $T_{\rm fit}$ is relatively straightforward given what we have learned in the previous section about how to alter the slope of $T_{\rm fit}$. 
We simply need to alternate between increasing and decreasing the harmonicity of the potential --- something that would naturally occur in the presence of breathing modes incited by a perturbation.
We keep the construction of the toy model the same, but we let $\omega=0.3$ inverse time units so that the period of oscillation is $\simeq 20$ time units.
Doing this means the $\ln \cosh$ contribution to the total potential now follows the purple curve in the top panel of \figref{toy_pot_change}.
As a result, $T_{\rm fit}$ oscillates over the course of the simulation (see bottom panel of \figref{toy_pot_change}), similar to the oscillations we see in \figref{tp_nbody_winding} for the N-body simulation.

\subsection{Critical assessment}
\label{sec:assess}

This section concentrated on exploring the extent to which deviations of phase-spiral evolution from simple phase-mixing could be explained by the expected properties and evolution of a perturbed and relaxing disk. We found that that the disturbance and subsequent re-alignment of the disk midplane could contribute to delays in the start of winding. We would expect this delay to be strongest in the inner parts of the galaxy where the disk dominates the gravitational potential. In addition, we found that the presence of breathing modes might cause the winding rate of the spiral to oscillate.

On the other hand, we were unable to attribute perhaps the most striking attribute of our simulated phase-spirals -- the slowing down of their winding -- to the expected evolution of the disk.  Rather, we found that any thickening of the disk would naturally speed up rather than slow down the rate of winding. In any case, no significant evolution of disk thickness was observed during the duration of the simulation.

This failure hints at the importance of dynamical effects that we have not yet explored. In particular, none of our toy models accounted for the existence of large-scale coherence in the disk response, time dependence of the halo potential, or coupling between the evolution of the two components. 
In addition, while our N-body simulation does not include gas or star formation, the real MW has both, which \citet{tepper2025galactic} have shown can influence phase spiral evolution.

\section{Discussion} \label{sec:discussion}

\subsection{Connecting simulation results to the Milky Way} \label{sec:mw}

In the left and right panels of \figref{freq_and_accel}, we make the same frequency and acceleration plots as for the N-body simulation, but based on the best MW potential currently available \citep[\texttt{MilkyWayPotential2022} from the Gala dynamics package;][]{Gala:2017}.
While there are some differences between the MW and our N-body simulation -- most notably that $\nu/\Omega$ is larger throughout the disk (see \figref{freq_and_accel}) -- we see similar trends to the N-body simulation, especially how the local contribution to the acceleration decreases with increasing $J_\phi$.
This means that the qualitative conclusions from our N-body simulation are likely to apply to the MW as well.

\begin{figure*}[t!]
    \centering
    \includegraphics[width=\linewidth]{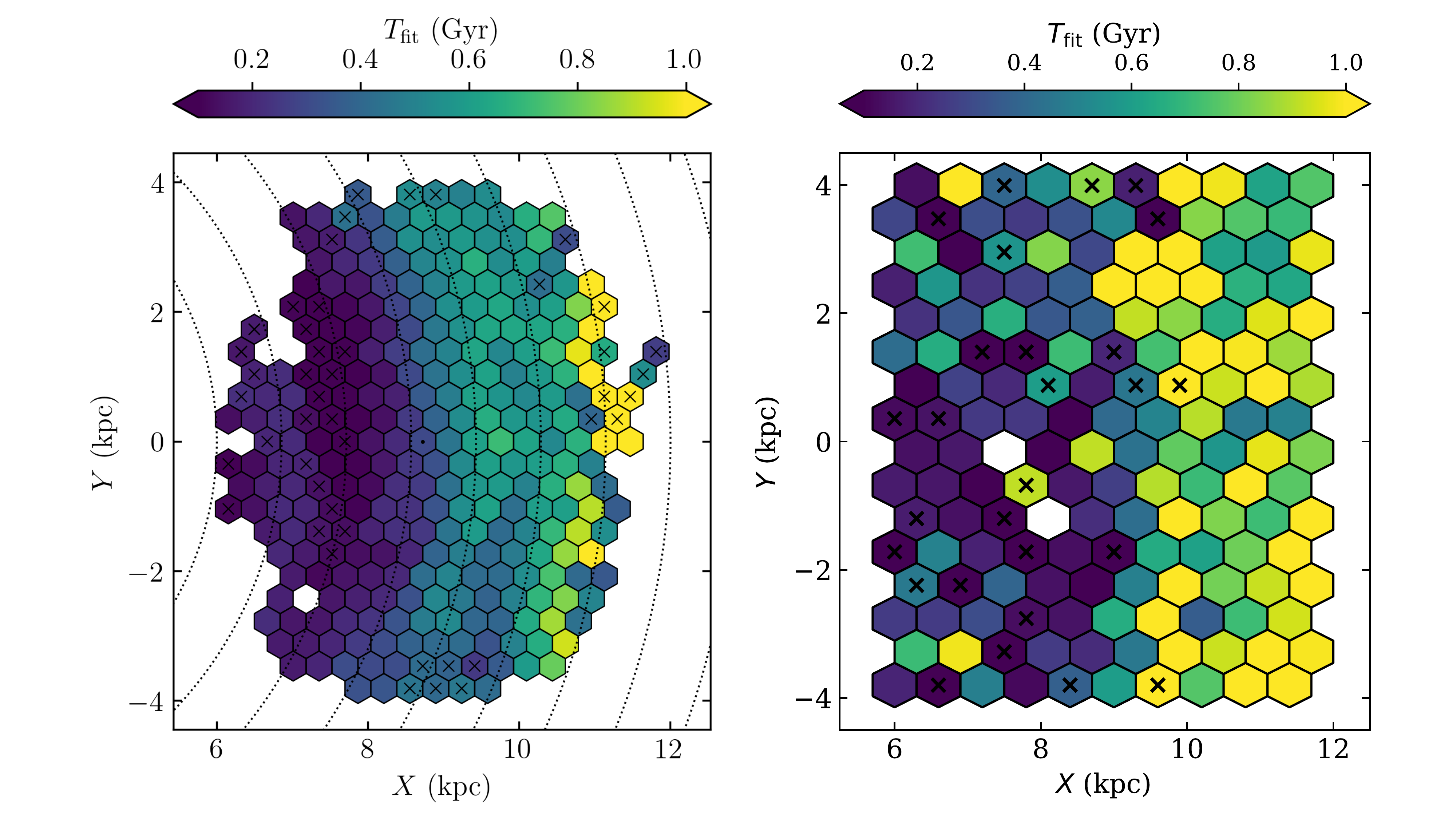}
    \caption{\textit{Left:} A combination of Figs. 4 and 9 from \citet{Widmark:2025} showing winding times for phase spirals across the MW disk using \emph{Gaia} DR3. 
    \textit{Right:} Same as the left panel but for an equivalent region in the N-body simulation at a timestep 800 \Myr after the disk crossing. The regions marked with an ``x'' are those where the derivation of $T_{\rm fit}$ might not be trustworthy, because $\theta_{z,{\rm max}}(J_z)$ is not monotonic. We must use larger bins than for the MW data because the simulation is lower resolution than \Gaia DR3.
    The two panels have notable similarities, most notably the increasing $T_\mathrm{fit}$ with increasing Galactocentric radius and the large-scale spatial coherence of $T_\mathrm{fit}$.}
\label{fig:mw_nbody_tfit}
\end{figure*}

We isolate a couple key points where we can compare our results to observations of the MW so far.
In our N-body simulation, at 1 Gyr after a single perturbation, (second panel of \figref{nbody_tfit}), we find that winding times vary by 100's of Myrs across the face of the disk.
Moreover, this variation is coherent, with a clear gradient as a function of $J_\phi$ and even some evidence for diagonal ridges.
Both of these findings are reminiscent of recent derivations of winding times in the MW \citep[\eg][]{Frankel:2023, Darragh-Ford:2023, Antoja:2023, Widmark:2025}. 
To show this, we combine Figs. 4 and 9 from \citet{Widmark:2025} to create the left panel of \figref{mw_nbody_tfit}.
This shows a clear trend of steadily increasing $T_{\rm fit}$ with increasing radius, as well as coherent large-scale structure in phase spiral morphology across the face of the disk.
To allow for the easiest comparison to the N-body simulation, we make the equivalent plot (hexagonal bins in $X-Y$) for the N-body simulation in the right panel of \figref{mw_nbody_tfit}, at a timestep 800 \Myr after the disk crossing.
The figure shows very similar trends to the data.
In both plots, the regions marked with an ``x'' are those where the derived $T_{\rm fit}$ values are less trustworthy. 
For the N-body case we determine the trustworthiness of a $T_{\rm fit}$ calculation based on how monotonic $\theta_{z,{\rm max}}(J_z)$ is.
We note that this is quite a conservative cut, so we choose to still show the derived $T_{\rm fit}$ values for these regions.

Based on our analysis, we can also offer a suggestion for how to find the true winding time from our observational results.
Our N-body simulation clearly demonstrates that, for a disk where phase spirals are induced by a satellite, $T_\mathrm{fit}$ calculated from phase spirals in the outer disk are more accurate than those near the Solar Neighborhood or in the inner disk.
Our toy models, where the perturbation is more generic, suggest that this is likely to be true for other types of perturbations as well.
This gives us a clear strategy for finding the true perturbation time responsible for the current phase spirals in the MW: trust the winding times in the outer disk more than those in the inner disk.
Applying this to our results in the left panel of \figref{mw_nbody_tfit}, we suggest that, if the MW phase phase spirals share a common origin, the perturbation which caused them likely occurred nearly $1 \Gyr$ ago.

\subsection{Comparison with previous work} \label{sec:comparison}

As discussed in \secref{intro}, accounting for full N-body interactions 
has a significant influence on phase spiral formation and evolution. 
However, it is only recently that researchers have been able to look at phase spirals in fully self-consistent simulations of galactic disks, as we do here.
As a result, it is worth comparing our findings on the phase spiral delay to two previous similar studies.

\citet{Bland-Hawthorn_Tepper-Garcia:2021} used a $10^8$ particle N-body simulation with a Sgr-like satellite perturber to examine the evolution of phase spirals.
They found that at Galactocentric radii approximately equal to the Sun's, phase spirals only became visible almost 500 \Myr after the disk crossing. 
This suggests significant delay in the onset of winding, although they did not perform quantitative measurements for $T_{\rm fit}$ so the precise value of this delay is uncertain.
They also claimed that phase spirals emerge at around the same time for many different Galactocentric radii, but exploring this delay was not the central focus of their analysis so they did not go into detail.
Still, their results are qualitatively similar to ours.

Our work shares the most similarities with \citet{Asano&Antoja:2025}, which also compared phase spiral winding in test particle and N-body simulations.
Their simulations are more similar to the true MW than ours, both in the construction of the host galaxy and when recreating the orbit of the Sgr satellite.
As in our work, they find that in the self-consistent N-body simulation, the phase spiral winding is delayed. 
Interestingly, however, they do not see the trend of decreasing delays at larger Galactocentric radii, and in fact suggest that the delay may be more significant in the outer disk.
This is perhaps partially explained by the fact that their analysis only covers $5\lesssim R_G \lesssim 13$ \kpc, which does cut out the critical $J_\phi \simeq 3000-4000 \kpckms$ range where we see the largest delay gradients.
Still, even in the overlapping range, neither our results nor recent observational analyses (see \secref{mw}) seem to agree with their trend, suggesting further explorations into this subject are needed.

One exploration included in \citet{Asano&Antoja:2025} that we do not replicate is looking at additional test particle simulations with varying components that make up a satellite flyby.
In particular, they separate the perturbation into the satellite itself and its dark matter wake, which \citet{Grand:2023} suggested as a possible cause for the Gaia phase spiral.
Doing so, they found that since the dynamical friction dark matter wake of the satellite crosses the disk after the main body does, this can create the semblance of a delay in the phase spiral winding. 
They show this delay can be up to 100 \Myr.
While important to consider, this phenomenon accounts neither for the extent of the delay that they find, nor the one in our work, meaning that the self-consistency of the simulation is still the crucial factor. 


\subsection{Implications for studying the Milky Way phase spirals' origins} \label{sec:origin}

While we cannot constrain the MW phase spirals' origins using our test particle and N-body simulations, we can contribute to the question of how we should interpret observational results.
As discussed in \secref{mw}, recent studies of the MW phase spirals have led to a large range of derived winding times ($0.2-1 \Gyr$) \citep[\eg][]{Frankel:2023, Darragh-Ford:2023, Antoja:2023, Widmark:2025, Hunt&Vasiliev:2025} with significant spatial coherence \citep[see the left panel of \figref{mw_nbody_tfit};][]{Widmark:2025}.
This has posed several conundrums when attempting to constrain the origin of the phase spirals.
On one hand, differences in winding times throughout the disk suggest perturbations at different times. 
At first glance, this appears to rule out origins from a single perturbative object like Sgr and was one of the catalysts for the stochastic small and large scale kicks proposal put forward by \citet{Tremaine:2023}.
On the other hand, the coherence of spiral properties (\eg pitch angle, phase angle, amplitude) across large regions of the disk suggests global perturbations, which are likely to be fewer in number. These contradicting results mean we have to question the interpretation of our observational evidence.

Our results contribute to solving this contradiction by showing that spatial coherence and winding time variations are compatible with a single perturbation as the source.
In particular, it is clear from our and prior work that variations in winding times across the disk can arise from a single perturbation due to: (i) differences of $\lesssim 100 \Myr$ when different regions of the disk experience the largest force from that perturbation; (ii) an evolving potential for the vertical structure of the disk; and (iii) differential mixing of signals across the disk.

\section{Summary and Conclusions} \label{sec:conclusion}

\subsection{Summary}

The 
dynamics of phase spiral formation and evolution remains an outstanding question in Galactic dynamics.
In this paper, we focus on developing a physical understanding of the dynamics at play rather than attempting to match the observations.
Specifically, we have concentrated on isolating the effect that incorporating full N-body interactions has on phase spiral formation and evolution. 

To do so, we adopted a new method of characterizing phase spirals: basis function expansions.
BFEs are a useful way to study phase spirals quantitatively because they make it easy to separate components with different numbers of spiral arms from both each other and background noise.
This allowed us to study the one-armed phase spirals in both a test particle and a self-gravitating simulation.

In the test particle case, the phase spirals wind up as expected from pure phase mixing theory.
Therefore, we were able to use the derived winding times ($T_\mathrm{fit}$) to accurately recover the disk crossing time of a Sgr-like satellite.
In fact, the winding time calculations were so precise that we could see a dipole in $T_\mathrm{fit}$ based on whether the region was more perturbed before or after the satellite passed through the disk (see \figref{tp_dipole}).
We performed this analysis on the test particle case to ensure that both our BFE reconstructions and winding time derivations were working accurately before applying both to the self-consistent simulation.

In the N-body simulation, the same analysis revealed significant deviations for $T_{\rm fit}$ derived from the phase-spirals compared to our expectations due to phase-mixing alone. These deviations were particularly pronounced in the inner galaxy where the disk dominates the gravitational potential.

We used toy models to explore the extent to which the expected evolution of the disk potential (e.g. oscillations of the midplane, breathing modes and disk thickening) could explain the deviations in $T_{\rm fit}$ that we observed. We found that the models provided useful intuition for some, but not all, of the behavior we could see.

\subsection{Conclusions}

Our results confirm prior work, which argued that the influence of N-body interactions must be considered in order to use winding times of phase-space spirals to deduce something about their origins. 
This work moves us towards being able to make that association with the following key results.
\begin{itemize}
    \item Relative to our expectations from pure phase mixing theory we find that in the N-body simulation: (i) the onset of winding is delayed; (ii) the rate of winding is slowed; and (iii) the rate of winding oscillates. Both (i) and (ii) become less important with increasing $J_\phi$ and are negligible for $J_\phi \gtrsim 3000 \kpckms$ ($R_G \gtrsim 14\kpc$).
    \item Some of these findings can be intuitively interpreted as a consequence of the phase-spirals evolving in a perturbed and relaxing disk potential. Our toy models show how the onset of winding can be delayed until the midplane is re-established; and the rate of winding can oscillate as the disk's vertical potential itself also oscillates due to (e.g.) breathing modes.
    \item Disk evolution alone does not provide a viable explanation for the slowing of the rate of winding in the inner disk compared to expectations. This failure suggests significant dynamical effects yet to be explored, such as the time evolution of the halo component, the coupling of the halo and disk, and the bulk motion of the disk towards the satellite as the latter approaches. 
    \item A map of the MW's winding times has similar trends and scales of variations as the N-body simulation 800 \Myr after the disk crossing (see \figref{mw_nbody_tfit}). Since these maps are dependent on the disk's response as opposed to the nature of the perturbation, it suggests that phase spirals in the MW are compatible with a single origin, such as Sgr.
\end{itemize}

Overall, our results demonstrate that it is critical to consider the way in which N-body interactions can affect phase spiral dynamics and that ignoring it will lead to inaccurate perturbation times.
Phase spiral morphologies and winding times can potentially be used to: distinguish formation scenarios (\eg Sgr, subhalos, dark matter halo distortions); and measure physical quantities (\eg orbit of Sgr, frequency of subhalo interactions, scale of dark matter halo distortion).
We conclude that, when using winding times for physical interpretations:
(i) winding times derived from MW phase spirals should be considered lower limits; and (ii) the winding times in the outer disk are likely to be more accurate than those in the inner disk or the Solar Neighborhood.

\vspace{6pt}
\section*{Acknowledgments}
We would like to thank the EXP collaboration and the Nearby Universe Group at the Center for Computational Astrophysics (CCA) for useful discussions and feedback during this project. K.T. also thanks the \textit{Winding, Unwinding, and Rewinding the Gaia Snail} workshop organizers and participants for useful conversations in August 2025.
K.V.J. is supported by Simons Foundation grant 1018465. J.A.S.H acknowledges the support of a UKRI Ernest Rutherford Fellowship ST/Z510245/1. A.W. is supported by the European Union's Horizon 2020 research and innovation program, under the Marie Skłodowska-Curie grant agreement number 101106028. C.H. is supported by the John N. Bahcall Fellowship Fund at the Institute for Advanced Study. M.S.P. acknowledges the support of a UKRI Stephen Hawking Fellowship. 
\bibliography{main}{}
\bibliographystyle{aasjournal}

\end{document}